\documentclass[preprint]{aastex}
\usepackage{graphics}
\usepackage{graphicx}
\usepackage[usenames]{color}

\newcommand\ie{{\em i.e.}\ }
\newcommand\eg{\textsl{eg.}\ }
\newcommand\etal{{\em et al.}\ }

\newcommand{\ly}{Lyman-$\alpha$ }

\begin{document}

\title{The Hot Inter-Galactic Medium and the Cosmic Microwave Background}

\author{Michael Fisher}
\affil{Department of Astronomy\\The Ohio State University\\140 West 18$^{\rm th}$ Avenue, Columbus, OH  43210-1173}
\affil{Current address:\\Battelle\\505 King Avenue, Columbus, OH  43201-2693}
\email{fisherml@battelle.org}

\begin{abstract}
The physical characteristics of the \ly forest cloud systems are combined with observations on the baryonic mass density of the Universe and constraints from primordial nucleosynthesis to set boundary conditions on the Intergalactic Medium (IGM) at the epoch of $z=2.5$. The Universe is considered a closed system and allowed to expand adiabatically from the epoch when QSOs first ionized the IGM ( $5 \leq z_{on} \leq 20$). The average kinetic energy of a gas is calculated in the region where the gas transitions from relativistic to non-relativistic behavior. All of the above measurements are then used to determine the thermal history of the IGM in the redshift range $2.5 \leq z \leq z_{on}$. The hot IGM is assumed to inverse Compton scatter photons from the Cosmic Microwave Background (CMBR) and consequently distort the CMBR as seen at the present epoch. The temperature of the IGM at $z=2.5$ and the epoch $z_{on}$ are adjusted, within the constraints defined above, to give the best overall agreement with published data on the temperature of the IGM. We find that the model of the IGM proposed here does not grossly distort the CMBR, and in fact agrees quite closely with the preliminary results from the Cosmic Background Explorer (COBE) satellite. However, our model of the IGM cannot explain the observed cosmic x ray background.

This paper was originally written in 1990. It was never submitted for publication.
\end{abstract}

\section{Introduction}

In the late 1940s a ``Big Bang'' theory of nucleosynthesis was postulated by George Gamow and his students, Ralph Alpher and Robert Herman. In 1948, Alpher and Herman used this theory to predict a blackbody background with a present temperature of about $5~K$. Similar calculations were done with similar results by Ya. B. Zeldovich, Fred Hoyle and R.J. Taylor in 1964. Independent of the above calculations, P.J.E. Peebles argued in 1965 that there ought to be a blackbody background of radio noise left over from the early Universe, red shifted to a present temperature of about $10~K$. Peebles based his argument on the fact that if there had not been an intense radiation field during the initial few minutes of the expanding Universe, that nucleosynthesis would have proceeded at such a rate that a substantial amount of the primordial hydrogen would have been fused into heavier elements. This would be in direct conflict with the fact that roughly 75 percent of the observed baryonic matter of the Universe is hydrogen. The intense radiation field was needed in order to prevent the hydrogen nuclei from fusing.

This radiation field would have survived, albeit at lower and lower temperatures as the Universe continued to expand. Therefore this radiation would be present today and if there had been no interaction of the radiation field with other constituents of the Universe, the radiation field would be isotropic, homogeneous and a perfect blackbody. In 1964, Arno Penzias and Robert Wilson discovered this relic radiation field using a 20 foot horn reflector antenna located on Crawford Hill at Holmdale, New Jersey.

Since 1965 a variety of experiments have been performed in order to confirm that Penzias and Wilson actually observed the relic radiation of the Big Bang. It is hypothesized that this relic radiation, with a temperature of $T_{\gamma 0} \approx 3~K$, would be dominate in the wavelength range $1~cm \geq \lambda \geq 0.3~mm$. The wavelength at which the maximum intensity would occur is found from the Wien displacement law
\begin{equation}
\lambda_{max} T_{\gamma} =0.29~~~.
\end{equation}
For $T_{\gamma} = 3~K$, $\lambda_{max} \approx 1~mm$. Unfortunately this entire range of wavelengths is not observable from the earth's surface. Molecular absorption bands in the earth's atmosphere prevent observations at wavelengths less than $1~cm$. Rocket borne instruments are needed to observe the Wien Tail of the relic radiation.

A number of groups have measured the intensity of the relic radiation and have found it to be remarkably isotropic (discounting the dipole moment observed because of our galaxy's peculiar motion in the direction of the Virgo cluster of galaxies), and well fit by a blackbody {\em except} in the Wien Tail of the spectrum, $\lambda < 1~mm$. This data is shown in Table~\ref{cmbr} below. As seen in Table~\ref{cmbr}, measurements of $T_{\gamma}$, with the exception of the COBE data, indicate a considerable short-wavelength excess in the cosmic microwave background radiation, CMBR. Many attempts were made to explain blackbody distortions of this type, including:
\begin{enumerate}
\item Very early superstars or hypercactive protogalaxies released extraordinary amounts of radiant energy that was then reradiated in the infrared by a prodigious density of cosmic dust.
\item A hot IGM would Compton scatter photons off of energetic electrons and thereby raise the photon energy.
\end{enumerate}
It should be noted that in both scenarios above, the parameters of the model have been adjusted in order to provide a ``best fit'' for the observations. The first model has great difficulty in providing the necessary energy and amount of dust to fit the observed excess. The second model is easier to formulate and match to the observations. However this model requires that all of the baryonic mass of the Universe be contained in the IGM, \eg Taylor and Wright (1989) (hereafter TW) use $\Omega_{IGM} =0.41$, with $H_0=50~km~s^{-1}~Mpc^{-1}$. This is in direct conflict with the constraint on $\Omega_{baryon}$ by primordial nucleosynthesis, Steigman (1989), $\Omega_{b}h_0^2 < 0.02$. The claim made by TW that primordial nucleosynthesis constraints are ``not in strong conflict with our results since extrapolation from current D (deuterium) abundance measurements back to primordial abundances is somewhat uncertain since it depends sensitively on the amount of stellar processing, and therefore on galactic and stellar evolution models.''

\begin{table}[t]
\begin{center}
\begin{tabular}{ccc}
\hline \hline
Wavelength ($cm$) & $T_{CMBR}~(K)$ & Reference \\
\hline
50.0 & $2.45 \pm 0.70$ & Sironi \etal (1987) \\
21.2 & $2.28 \pm 0.39$ & Levin \etal (1988) \\
12.0 & $2.79 \pm 0.15$ & Sironi \& Bonelli (1986) \\
8.1 & $2.58 \pm 0.13$ & DeAmici \etal (1988) \\
6.3 & $2.70 \pm 0.07$ & Mandolesi \etal (1986) \\
3.0 & $2.61 \pm 0.07$ & Kogut \etal (1988) \\
0.909 & $2.81 \pm 0.12$ & Smoot \etal (1985) \\
0.333 & $2.60 \pm 0.10$ & Johnson \& Wilkinson (1987) \\
1.2 & $2.783 \pm 0.025$ & Crane \etal (1989) \\
0.264 & $2.796^{+0.014}_{-0.039}$ & Crane \etal (1989) \\
0.132 & $2.75^{+0.24}_{-0.29}$ & Crane \etal (1986) \\
0.264 & $2.70 \pm 0.04$ & Meyer \& Jura (1985) \\
0.22 & $2.60 \pm 0.09$ & Bersanelli \etal (1989) \\
0.132 & $2.76 \pm 0.20$ & Bersanelli \etal (1989) \\
0.116 & $2.799 \pm 0.018$ & Matsumoto \etal (1988) \\
0.071 & $2.963 \pm 0.017$ & Matsumoto \etal (1988) \\
0.048 & $3.150 \pm 0.026$ & Matsumoto \etal (1988) \\
0.351 & $2.80 \pm 0.16$ & Peterson \etal (1985) \\
0.198 & $2.95^{+0.11}_{-0.12}$ & Peterson \etal (1985) \\
0.148 & $2.92 \pm 0.10$ & Peterson \etal (1985) \\
0.114 & $2.65^{+0.09}_{-0.10}$ & Peterson \etal (1985) \\
0.100 & $2.55^{+0.14}_{-0.18}$ & Peterson \etal (1985) \\
1.0 -- 0.05 & $2.735\pm 0.06$ & Mather \etal (1990) (COBE90)\\
\hline
\end{tabular}
\end{center}
\caption{Recent measurements of the CMBR.}
\label{cmbr}
\end{table}

Of course, results from the Cosmic Background Explorer (COBE), Mather \etal (1990), hereafter COBE90, seem to indicate that there is no excess in the Wien Tail of the CMBR. To quote one astronomer from the AAS meeting of January 1990 (when preliminary COBE results were announced), ``There is no hot inter-galactic medium.'' COBE90 is a bit less definitive stating only that the hot smooth IGM suggested to explain the cosmic X-ray background can be ruled out. The COBE results that support this claim originate from nine minutes of observation near the North Galactic Pole by the Far Infrared Absolute Spectrometer, FIRAS. The spectral measurements, from $1~cm$ to $0.5~mm$, show less than a one percent deviation from an ideal blackbody with equivalent temperature $2.735 \pm 0.06~K$. Although the COBE data certainly refutes the findings of the Nagoya-Berkeley Group, Matsumoto \etal (1988), the limited wavelength range of observations do not conclusively rule out the existence of a hot IGM.

\section{Implications of Primordial Nucleosynthesis for the Physical Conditions in Lyman-$\alpha$ Forest Clouds}

The hot big bang model of cosmology, the ``standard'' model, is now recognized as the most accurate description of the Universe's beginning and evolution. Since George Gamow and his collaborators put forth the idea of a hot ``Big Bang'' nucleosynthesis model in the late 1940s (Gamow (1946); Alpher \etal (1948); and Alpher and Herman (1950)), many authors have extensively studied primordial nucleosynthesis. Appreciable amounts of D and He$^3$ are not formed in stellar nucleosynthesis and are destroyed by thermonuclear reactions at temperatures above $10^6$ and $10^7~K$, respectively. But D and He$^3$ are observable and their abundances cannot be explained by non-exotic methods. However conditions are present during the first few minutes of a hot expanding Universe to create D and He$^3$. Nucleosynthesis began when the temperature of the primitive Universe had cooled to about $10^9~K$. This occurred approximately 100 seconds after the Big Bang. Protons fused with neutrons to form deuterium nuclei and deuterium nuclei could then fuse to form helium. Most of the helium in the Universe was formed at this time, along with deuterium and lithium. As the Universe continued to expand and cool, the temperature fell below that which is needed to sustain these reactions, and the fusion of D and He$^3$ stopped.

The number density of the fusion by-products and the rates at which various fusion reactions proceed must be known in order to set the limits on when Big Bang nucleosynthesis began and ended. This is not a difficult task as these rates are measured in the laboratory. Then the evolution of a gas of neutrons and protons, \ie baryons, is followed in time  to calculate the results of Big Bang nucleosynthesis. Before nucleosynthesis began, the protons and neutrons were in equilibrium and were present in equal numbers. Neutrons could change into protons and vice versa via collisions with members of the lepton family, electrons and neutrinos and their associated anti-particles. The neutron to proton rate is slightly larger than the proton to neutron rate because the neutron is slightly heavier than the proton. However, early in the Big Bang, these reactions proceed at nearly the same rate because of the high temperature.

The ratio of neutrons to protons fell from one to less than a third as the temperature fell to $10^7~K$ (because of the different reaction rates). At a temperature of $10^6~K$, this ratio fell below a seventh, and the temperature was sufficiently low to allow fusion to begin. The first fusion product was deuterium, one proton and one neutron in the nucleus. Interaction of deuterium with other protons and neutrons formed tritium (one proton and two neutrons) and He$^3$ (two protons and one neutron). These nuclei continued to interact to form He$^4$. He$^4$ is a very stable and tightly bound nucleus, consequently almost all of the neutrons were bound into He$^4$ nuclei, with traces left in D, He$^3$, Be$^7$ and Li$^7$. Nucleosynthesis ceased at He$^4$ since no stable nuclei is formed from the interaction of He$^4$ and a proton or a neutron.

The synthesis of these light elements depends only on one adjustable parameter, the density of nucleons. A higher nucleon density would ``burn'' D at a faster rate and hence the D abundance would be lowered. Deuterium is only destroyed in the course of galactic chemical evolution, therefore an upper limit can be set for the nucleon density from the local observed D abundance, Steigman (1989). D is fused by stars into He$^3$, which some survives stellar processing. In this manner, the local He$^3$ and D abundances can be used to set a lower bound on the nucleon density. These bounds can be refined by using the other light elements formed during primordial nucleosynthesis. Steigman (1989) finds that the present density of nucleons is bounded by
\begin{equation}
0.011 \leq \Omega_b h_{100}^2 \leq 0.020~~~,
\end{equation}
where $\Omega_b$ is the ratio of baryon mass to the critical mass of the Universe and $h_{100}^2$ is the Hubble constant in units of $100~km~s^{-1}~Mpc^{-1}$.

\section {Properties of \ly Clouds}

The ``\ly forest'' of absorption lines seen in the spectra of high redshift Quasi-Stellar Objects (QSOs) have been the subject of much investigation in recent years. Sargent \etal (1980) concluded that this forest of absorption lines resulted from discrete clouds of ionized gas along the line-of-sight and are not associated with the QSOs. Therefore these cloud systems can be used as windows for examining the Inter-Galactic-Medium as well as other cosmologically significant properties.

Typical physical characteristics for a \ly cloud are: size, $D \approx 15~{\rm kpc}$; temperature, $T_e \approx 40,000~{\rm K}$; and neutral hydrogen column density, $N_{HI} \approx 10^{15}~{\rm cm^{-2}}$. The clouds are assumed to be photoionized by the cosmic background with a total hydrogen density $n_{H}^{tot} \approx 10^{-3}-10^{-4}~{\rm cm^{-3}}$ and in pressure equilibrium with a confining hot phase. Several authors have also noted that the number of clouds per unit redshift increase with redshift and decrease with column density described by a distribution function of the form
\begin{equation}
\label{prob}
P(N_{HI},z)dN_{HI}dz = BN_{HI}^{-\beta}(1+z)^{\gamma} dN_{HI}dz~~~,
\end{equation}
where $\beta$, $\gamma$ and $B$ are constants to be evaluated from observations.The expression is valid over a range of column densities $10^{13}~{\rm cm^{-2}} \leq N_{HI} \leq 10^{16}~{\rm cm^{-2}}$, Hunstead (1988).Many authors have measured $\beta$ and $\gamma$ as shown in Table~\ref{beta}. The constant $B$ is then evaluated from observations and shown in Table~\ref{bb}. This work uses $\beta = 1.7 \pm 0.2$, $\gamma = 2.3 \pm 0.4$ and $B=1.53 \times 10^{10}$ and all calculations will be done at redshift $z=2.5$. Therefore Eq.~(\ref{prob}) can be written as
\begin{equation}
P(N_{HI},z)dN_{HI}dz = 1.53 \times 10^{10} N_{HI}^{-1.7} (1+z)^{2.3} dN_{HI}dz~~~,
\end{equation}
and the total neutral hydrogen density per unit redshift is
\begin{equation}
\frac{d}{dz} N_{HI} = \int N_{HI}P(N_{HI},z)dN_{HI} = 5 \times 10^{16} A_5~{\rm cm^{-2}}~~~,
\end{equation}
where $A_5$ is the uncertainty in the measured neutral hydrogen column density in units of $5 \times 10^{16}~{\rm cm^{-2}}$.

\begin{table}[t]
\begin{center}
\begin{tabular}{ccl}
\hline \hline
$\beta$ & $\gamma$ & Reference \\
\hline
$1.89 \pm 0.14$ & $1.7 \pm 1.0$ & Atwood \etal (1985) \\
$1.61 \pm 0.1$ & & Tytler (1987) \\
$1.68 \pm 0.10$ & & Carswell \etal (1984) \\
$1.71$ & & Carswell \etal (1987) \\
$1.65 \pm 0.10$ & & Hunstead \etal (1987) \\
& $2.36 \pm 0.36$ & Peterson (1983b) \\
& $2.20 \pm 0.40$ & Peterson (1983a) \\
& $2.17 \pm 0.36$ & Murdoch \etal (1986b) \\
& $1.40 \pm 0.70$ & Carswell \etal (1982) \\
\hline
\end{tabular}
\caption{Recent measurements of $\beta$ and $\gamma$.}
\label{beta}
\end{center}
\end{table}

\begin{table}[t]
\begin{center}
\begin{tabular}{ccl}
\hline \hline
$B \times 10^{10}$ & Method & Reference \\
\hline
1.53 & Total (high $z$) & Carswell (1988), Fig. 4 \\
0.63 & Statistical & Carswell (1988), Fig. 4 \\
3.16 & Total (low $z$) & Carswell (1988), Fig. 4 \\
0.01 & Statistical & Carswell (1988), Fig. 4 \\
2.17 & Total & Hunstead (1988), Fig. 4 \\
0.08 & Statistical & Hunstead (1988), Fig. 4 \\
0.17 & Total & Atwood \etal (1985) \\
0.398 & Statistical & Tytler (1987) \\
10.6 & Statistical & Carswell \etal (1984) \\
0.855 & Total & Carswell \etal (1987) \\
0.128 & Statistical & Carswell \etal (1987) \\
\hline
\end{tabular}
\caption{Determination of the constant $B$. Total method is normalized to neutral hydrogen column density and statistical method is normalized to linear least squares regression.}
\label{bb}
\end{center}
\end{table}

Observations cannot reveal the {\em ionized} column density, $N_{HII}$, through a cloud, only the {\em neutral} column density, $N_{HI}$. However, the former is needed to estimate the mass of a cloud since the cloud is mostly ionized. The ionized and neutral hydrogen number densities are related through a simple ionization balance equation
\begin{equation}
\label{balance}
n_e n_p \alpha(T_e) = \Gamma(z) n_{HI} ~~~,
\end{equation}
where $\alpha$ and $\Gamma$ are the recombination coefficient and photoionization rate, respectively. The clouds are much larger than the angular diameter of the QSO. Therefore the average chord length through a cloud along the line-of-sight is $\langle D \rangle = D/\sqrt{2}$. This suggests particle densities $n_{HI} = 3.05 \times 10^{-8} D_{15}^{-1}~{\rm cm^{-3}}$ at $z=2.5$, and $D_{15}^{-1}$ is the uncertainty in cloud diameter in units of $10^{15}~{\rm kpc}$. The ionization balance equation is normalized by assuming the clouds have primordial abundances, 90\% hydrogen and 10\% helium by number; helium is fully ionized, $n_e=1.2 n_p$; and a typical temperature of $T_e = 40,000~{\rm K}$, with corresponding Case B recombination coefficient, $\alpha_B(T_e)=7.7 \times 10^{-14}~{\rm cm^3~s^{-1}}$. The photoionization rate is defined as
\begin{equation}
\label{gamma}
\Gamma(z) = \int_{\nu_0}^{\infty} \frac{J_{\nu}}{h \nu} \sigma_{\nu} d \nu~~~.
\end{equation}

Bajtlik \etal (1988) solve for the background intensity from the proximity effect found in \ly forest lines. Their measured intensity is $\log J_{\nu}^{LL} = -21.0 \pm 0.5$, where $J_{\nu}^{LL}$ is the Lyman-limit intensity in units of ergs cm$^{-2}$ s$^{-1}$ Hz$^{-1}$. Using this form for the intensity, Eq. (\ref{gamma}) is then
\begin{equation}
\Gamma(z) = 2.99 \times 10^{-12}~J_{21}~{\rm s^{-1}}~~~,
\end{equation}
where $J_{21}$ is the uncertainty in the Lyman-limit diffuse background  in units of $4 \pi \times 10^{-21}~{\rm ergs~cm^{-2}~s^{-1}~Hz^{-1}}$. Eq. (\ref{balance}) can then be solved for the proton number density
\begin{equation}
n_p = 9.93 \times 10^{-4}~J_{21}^{1/2}~D_{15}^{-1/2}~{\rm cm^{-3}}~~~.
\end{equation}
Assuming the clouds retain their integrity, expand isothermally with the Universe and are in pressure equilibrium with a hot phase which is nonrelativistic (at $z=2.5$) and expanding adiabatically, \ie $P(z) = n_{tot}(z) T(z) \sim (1+z)^5$, the electron to neutral hydrogen density ratio is
\begin{equation}
\frac{n_e}{n_{HI}} = 3.91 \times 10^4~J_{21}^{1/2}~D_{15}^{-1/2}~~~.
\end{equation}
The electron column density per unit redshift is then
\begin{eqnarray}
\label{clouds}
\frac{d}{dz} N_e & = & \frac{d}{dz}N_{HI} \frac{n_e}{n_{HI}} ~~~, \nonumber \\
& = & 1.96 \times 10^{21} ~J_{21}^{1/2}~D_{15}^{-1/2}~A_5~{\rm cm^{-2}}~~~.
\end{eqnarray}

\section{Baryonic Mass Budget of the Universe at $z=2.5$}

The mass density contained within the \ly clouds can be determined from Eq. (\ref{clouds}). In an Einstein-de Sitter Universe
\begin{equation}
\frac{dl}{dz} = \frac{c}{H_0} (1+z)^{-5/2} ~~~.
\end{equation}
Since the clouds are of primordial abundances, the number density of any one cloud is then
\begin{eqnarray}
\rho_{{\rm cloud}} & = & m_p n_H + m_{\alpha} n_{He} ~~~, \nonumber \\
& = & 1.2 m_p n_e~~~.
\end{eqnarray}
The mass density of all of the clouds at redshift $z=2.5$ is then the mass of one cloud $\times$ the number density of clouds
\begin{equation}
\rho_{{\rm clouds}} = 1.2 m_p \frac{dN_e}{dz} \frac{dz}{dl} ~~~.
\end{equation}
This is evaluated to be
\begin{equation}
\rho_{{\rm clouds}} = 9.72 \times 10^{-30}~ h_{100}~J_{21}^{1/2}~D_{15}^{-1/2}~A_5~{\rm g~cm^{-3}}~~~.
\end{equation}
This result was used to determine the baryonic mass budget of the Lyman-$\alpha$ clouds
\begin{equation}
\Omega_{clouds} = \frac{\rho_{clouds}}{\rho_{critical}} = 0.012 h_{100}^{-1} J_{21}^{1/2} D_{15}^{1/2} A_5~~~.
\end{equation}
This result can be used to determine the conditions in the Inter-Galactic Medium at the epoch $z=2.5$.

\section{Density and Temperature of the IGM at the Epoch $z=2.5$}

The baryonic mass budget of the Universe remains an open question. Studies of primordial nucleo synthesis, Steigman (1989), show that the fraction of the critical density contained in baryons is
\begin{equation}
\Omega_b = 0.02~h_{100}^{-2} ~~~,
\end{equation}
where $h_{100}$ is the uncertainty in the Hubble parameter, $H_0 = 100 \times h_{100}~{\rm km~s^{-1}~Mpc^{-1}}$. It would be surprising if galaxy formation were 100\% efficient, and in fact stars and other luminous material in galaxies contribute $\Omega_{\rm galaxies} \approx 0.007-0.017$ at the present epoch, Tang \etal (1986) and Faber \& Gallagher (1979), \ie 19\% to 47\% of $\Omega_b$, for $h_{100} = 0.75$. This suggests that the majority of the baryonic mass might occur in another form, perhaps in the IGM. Results from the previous section indicate that the baryonic mass contained in the \ly clouds is
\begin{equation}
\Omega_{{\rm clouds}} = 0.6 h_{100} \Omega_b ~~~.
\end{equation}
The \ly clouds have been extensively studied. Peterson (1978) and Sargent \etal (1980) have shown that these clouds: are not associated with galaxies as evidenced by the lack of clustering, have low metal abundance and exist in large numbers. They are inter-galactic in nature and of primordial abundances. They are thought to be confined by pressure equilibrium with the rest of the IGM, $n_{{\rm IGM}} T_{{\rm IGM}} = n_{{\rm cloud}} T_{{\rm cloud}}$. We now assume that the remainder of the baryonic mass, as indicated by primordial nucleosynthesis, is contained in the rest of the IGM, $\Omega_b = \Omega_{{\rm galaxies}} + \Omega_{{\rm clouds}} + \Omega_{{\rm IGM}}$. Since the IGM is also of primordial abundance, the number density of the IGM is
\begin{eqnarray}
\Omega_{{\rm IGM}} & = & \Omega_b (1-0.35 h_{100}^2 - 0.60 h_{100})~~~, \\
\rho_{{\rm IGM}} & = & \rho_{{\rm critical}} \times \Omega_{{\rm IGM}} ~~~, \\
n_{{\rm IGM}} & = & \frac{\rho_{{\rm IGM}}}{1.2m_p} ~~~.
\end{eqnarray}
The above equations can be evaluated for $n_{{\rm IGM}}$
\begin{equation}
n_{{\rm IGM}} = 7.42 \times 10^{-6} (1-0.35 h_{100}^2 -0.60 h_{100}) ~~~.
\end{equation}

A lower limit can be set on the temperature of the IGM, at $z=2.5$, by choosing the minimum mass in the luminous matter, $\Omega_g$ and assuming pressure equilibrium with the \ly clouds
\begin{equation}
T_{{\rm IGM}} \geq \frac{5.35 \times 10^6}{1-0.35 h_{100}^2 -0.60 h_{100}}~{\rm K}~~~.
\end{equation}
With $h_{100}$, $T_{{\rm IGM}} \geq 1.5 \times 10^7~{\rm K}$ at $z=2.5$.  The lack of observed neutral hydrogen, Gunn \& Peterson (1965), indicates that if the IGM exists, beyond \ly systems, it is ionized and hot as indicated by the above temperature relationship.

\section{Thermal History of the IGM}

We now seek solutions to the temperature-redshift relation for the IGM. The temperature of the IGM at any epoch will directly affect the interaction of the cosmic microwave background radiation (CMBR) with the hot electrons of the IGM. We have already established the temperature of the IGM at the epoch $z=2.5$, and this will be used as a boundary condition for the temperature-redshift relation as a function of redshift.

Consider the Universe as a closed system and any expansion or contraction is consequently adiabatic. Therefore solutions with constant entropy are desired for the thermal history of the IGM. The entropy per particle of a non-relativistic ideal gas is
\begin{equation}
\frac{S}{k} = \frac{3}{2} \ln T + \ln V ~~~,
\end{equation}
where $k$ is Boltzman's constant, $T$ is the temperature of the gas and $V$ is the volume of the gas. Non-relativistic expansion indicates that $T \sim V^{-2/3}$, and since the volume scales as $V \sim (1+z)^{-3}$, the temperature scales as $T \sim (1+z)^2$, and the case of cooling a non-relativistic gas by adiabatic expansion is solved, Field \& Perrenod (1977), hereafter FP.

Unfortunately for this result, the IGM electrons are already nearly relativistic at $z=2.5$ as determined previously, and will become relativistic as the Universe contracts and the temperature increases. Therefore the cooling of a relativistic gas by adiabatic expansion must also be considered. A dimensionless temperature, $\theta \equiv kT/m_e c^2$ is defined such that $\theta = 1$ for $T=5.94 \times 10^9$ K. The general solution for the entropy per particle of any gas is given by
\begin{equation}
\label{electron}
\frac{S}{k} = \int_{\theta_0}^{\theta} \frac{dE}{d \theta} \frac{d \theta}{\theta} + \ln V ~~~,
\end{equation}
where the energy is expressed in units of $m_e c^2$. For a purely relativistic gas, $E = 3 \theta$, $dE/d \theta$ = 3, and the entropy per particle is
\begin{equation}
\frac{S}{k} = 3 \ln \left( \frac{\theta}{\theta_0} \right) + \ln V ~~~.
\end{equation}
Therefore in an adiabatic expansion, $\theta \sim V^{-1/3} \sim (1+z)$. The IGM consists of electrons , protons and $\alpha$-particles. The protons and $\alpha$-particles will always be non-relativistic for the temperatures considered here, but the electrons may become relativistic. The entropy per particle for this gaseous mixture is then
\begin{equation}
\frac{S}{k} = \int_{\theta_0}^{\theta} \frac{dE}{d \theta} \frac{d \theta}{\theta} + \ln V + \frac{10}{11} \left[ \frac{3}{2} \ln \left( \frac{\theta}{\theta_0} \right) + \ln V \right] ~~~.
\end{equation}
In the non-relativistic limit, $E = 3kT/2$, and in the relativistic limit, $E=3kT$. The dependence of the electron energy upon temperature in the transition region between these two limits must be known to solve Eq. (\ref{electron}).

\begin{figure}[t]
\begin{center}
\includegraphics{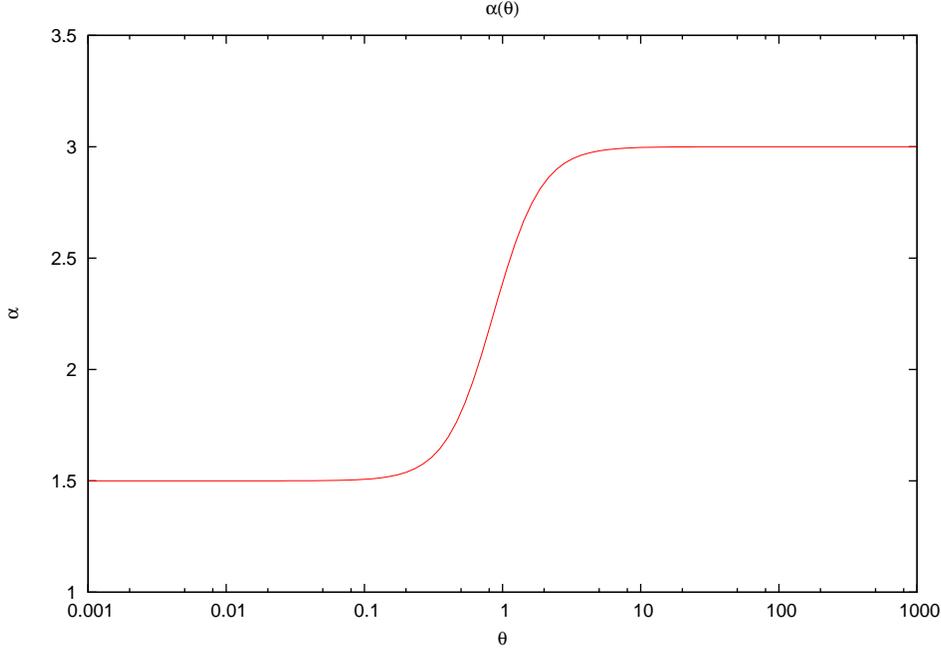}
\caption{The function $\alpha(\theta)$. Note that at low $\theta$, \ie non-relativistic temperatures, $\alpha = 3/2$ and the average kinetic energy is $3kT/2$. For large $\theta$, \ie relativistic temperatures, $\alpha = 3$ and the average kinetic energy is $3kT$. The transition between the two limits is smooth, ``nature abhors sharp corners,'' Professor of Physics Emeritus Stan Christensen, Kent State University.}
\label{alphafig}
\end{center}
\end{figure}

Consider any quantity $A$ of a system of particles, of mass $m$, that is dependent upon the momentum of the particles. Then the average value of $A$, $\langle A \rangle$, Tolman (1917), is
\begin{equation}
\langle A \rangle = 4 \pi V \int_0^{\infty} a \exp \left(\frac{c}{kT}\sqrt{\psi^2 + m^2 c^2} \right) A \psi^2 d \psi ~~~,
\end{equation}
where $a$ is a constant and $\psi$ is the momentum. The energy, in units of $m_e c^2$ is then
\begin{equation}
\label{energy}
\langle E \rangle = \frac{4 \pi V}{mc} \int_0^{\infty} a \exp \left(\frac{c}{kT}\sqrt{\psi^2 + m^2 c^2} \right) \sqrt{\psi^2 + m^2 c^2} \psi^2 d \psi ~~~.
\end{equation}
The unknown constant $a$ may be eliminated by normalization
\begin{equation}
4 \pi V \int_0^{\infty} a \exp \left(\frac{c}{kT}\sqrt{\psi^2 + m^2 c^2} \right) \psi^2 d \psi = 1~~~.
\end{equation}
The substitution
\begin{equation}
\gamma = \sqrt{1 + \frac{\psi^2}{m^2 c^2}} ~~~,
\end{equation}
simplifies Eq. (\ref{energy}) and the average kinetic energy per particle of the electron gas becomes
\begin{equation}
\label{thisisit}
\langle E \rangle = \frac{\int_1^{\infty} e^{-\gamma/\theta} \gamma^2 \sqrt{\gamma^2 - 1} d \gamma}{\int_1^{\infty} e^{-\gamma/\theta} \gamma \sqrt{\gamma^2 -1}} -1 ~~~,
\end{equation}
which corresponds to Eqs. (8) and (9) of TW. TW solved the denominator of Eq. (\ref{thisisit}) by integrating in parts to achieve $\theta K_2(1/\theta)$, where $K_2$ is a modified Bessel function of the second kind. To complete the integral in the numerator of Eq. (\ref{thisisit}), we note from the recursion relations for modified Bessel functions, Abramowitz \& Stegun (1964),
\begin{equation}
\frac{d}{dz} K_z(z) = \frac{2}{z} K_2(z)-K_3(z) ~~~.
\end{equation}
Integration by parts with the above relation, the numerator of Eq. (\ref{thisisit}) is
\begin{equation}
\int_1^{\infty} e^{-\gamma/\theta} \gamma^2 \sqrt{\gamma^2 -1} d \gamma = \theta K_3(1/\theta) - \theta^2 K_2(1/\theta) ~~~,
\end{equation}
and the average kinetic energy is
\begin{equation}
\langle E \rangle = \theta \left[ \frac{1}{\theta} \left( \frac{K_3(1/\theta)}{K_2(1/\theta)} -1 \right) -1 \right] = \alpha(\theta) \theta ~~~.
\end{equation}
The function $\alpha(\theta)$ is shown in Figure \ref{alphafig}.
The derivative of energy with respect to temperature (needed for the entropy per particle calculation) is
\begin{equation}
\label{deriv}
\frac{dE}{d \theta} = \frac{1}{\theta^2} - \left( \frac{K_3(1/\theta)}{\theta K_2(1/\theta)} \right)^2 + \frac{K_3(1/\theta)}{\theta K_2(1/\theta)} - 1 ~~~.
\end{equation}
The volume may be eliminated from Eq. (\ref{electron}) by $V = V_0 (1+z)^{-3}$. Then Eq. (\ref{deriv}) is used to integrate Eq. (\ref{electron}) to define the entropy per particle for both relativistic and non-relativistic adiabatic cooling by expansion. An additional source of cooling for the electron gas in the IGM is from inverse Compton scattering of CMBR photons off of hot IGM electrons. The change in entropy with respect to redshift from inverse Compton scattering is
\begin{equation}
\frac{1}{k}\frac{d S_{cs}}{dz} = \frac{1}{kT} \frac{dQ}{dz} = \frac{1}{\theta} \frac{dE}{dz} ~~~.
\end{equation}
The Compton power loss per scattering, Rybicki \& Lightman (1979), is
\begin{equation}
\frac{dE}{dt} = \frac{4}{3} \sigma_T c \gamma^2 \beta^2 \frac{U_{ph}}{m_e c^2}~~~,
\end{equation}
where $\sigma_T$ is the Thompson scattering cross section, $\beta$ has its usual relativistic meanings, $\gamma$ is as defined before, and $U_{ph}$ is the initial photon energy density, in this case the undisturbed CMBR, $U_{ph} = a T_{\gamma}^4$, where $T_{\gamma}$ is the photon temperature. Therefore we can write the entropy change for one electron due to inverse Compton scattering as
\begin{equation}
\frac{1}{k} \frac{d S_{cs}}{dz} = \frac{1}{\theta} \left( \frac{dE}{dt} \frac{dt}{dz} \right) ~~~.
\end{equation}

Setting $\Omega_0 = 1$ implies $dt/dz = -H_0^{-1} (1+z)^{-5/2}$. The power loss per electron is averaged over the electron distribution
\begin{equation}
\frac{1}{k} \frac{dS_{cs}}{dz} = - \frac{1}{\theta} \int_1^{\infty} \frac{4 c \sigma_T a T_{\gamma 0}^4}{3 H_0 m_e c^2} (1+z)^{3/2} \beta^2 \gamma^2 p(\gamma) d \gamma ~~~,
\end{equation}
where the substitution $T_{\gamma} = T_{\gamma 0} (1+z)$, as the CMBR photons are always relativistic. The integral is evaluated to be
\begin{equation}
\frac{1}{k} \frac{dS_{cs}}{dz} = - \frac{4 c \sigma_T a T_{\gamma 0}^4}{H_0 m_e c^2} (1+z)^{3/2} \frac{K_3(1/\theta)}{K_2(1/\theta)} ~~~.
\end{equation}

\begin{figure}[t]
\begin{center}
\includegraphics{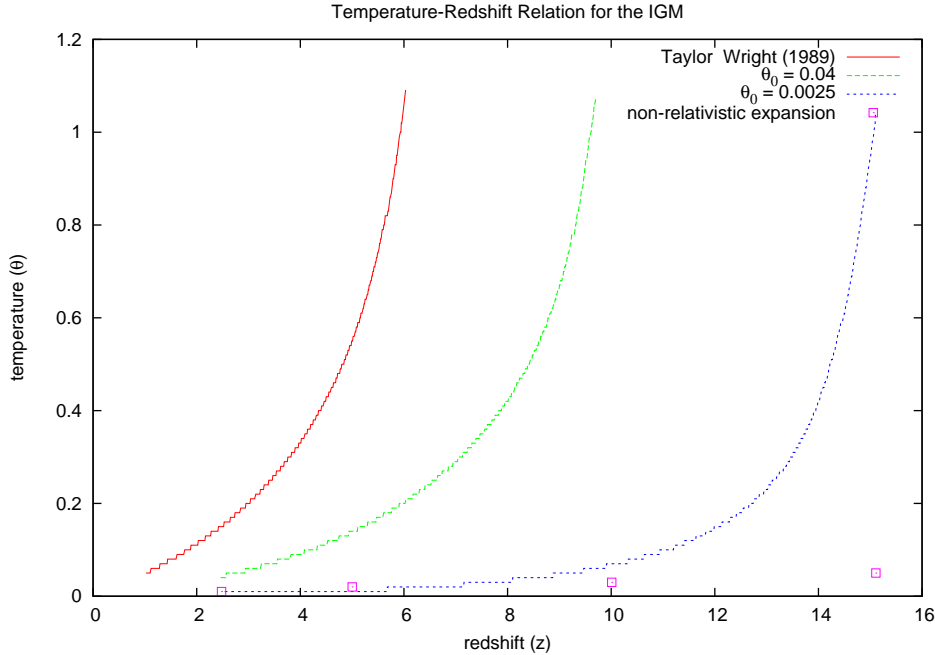}
\caption{The thermal history of the IGM for $H_0 = 75~{\rm km~s^{-1}~ Mpc^{-1}}$, for different initial temperatures $\theta$. Also shown is the model proposed by TW and the temperature from purely non-relativistic adiabatic expansion.}
\label{history}
\end{center}
\end{figure}

The total entropy per particle of the IGM is then
\begin{equation}
\label{entropy}
\frac{S}{k} = \int_{\theta_0}^{\theta} \frac{dE}{d \theta} \frac{d \theta}{\theta} + \frac{15}{11} \ln \left( \frac{\theta}{\theta_0} \right) - \frac{63}{11} \ln \left( \frac{1+z}{1+z_0} \right) + \int_{z_0}^z \frac{1}{k} \frac{dS_{cs}}{dz} dz ~~~.
\end{equation}
The temperature $\theta$ required to hold Eq. (\ref{entropy}) constant is found as a function of redshift. This defines the temperature-redshift relationship for the IGM, Figure \ref{history}. Also shown in Figure \ref{history} is the case for non-relativistic expansion and the model calculated by TW. The boundary conditions in Figure \ref{history} are $\theta = 0.00252$ at $z=2.5$, as found previously (lower limit), and $\theta =0.04$ as an intermediate value. The difference between this work and TW is evident in Figure \ref{history}. The differences result from the boundary conditions. TW chose $\theta_0$ in order to fit the observed x ray background. This work calculates $\theta_0$ from pressure equilibrium between the \ly cluds and the constraints on the fractional mass contained in baryons by primordial nucleosynthesis. It is clearly evident that the model proposed by TW would have a substantially greater effect upon the CMBR than the model presented here.

\section{Distortion of the CMBR by the Hot IGM}

We have previously shown that the IGM, if it exists, is hot and optically thin. Inverse Compton scattering, an efficient cooling mechanism for the IGM, also redistributes energy from the hot electrons to the photons of the CMBR. The redistribution of energy will in turn change the distribution in frequency of the photon gas, hence distorting the blackbody spectrum of the relic radiation. The effect is most noticeable in the Wien tail of the spectrum, and many authors have previously calculated this distortion in efforts to understand the reported excess in the Wien tail of the CMBR. Therefore this work is not unique in method, but because of the vastly different thermal history of the IGM presented in the previous section as compared to other author's efforts (most notable TW and FP), our calculated distorted CMBR is correspondingly different, albeit of the same nature as TW and FP.

The reader is directed to TW for a superb description of the methodology for calculating the distortion of the CMBR, it will not be reproduced here. Only the results will be reported.

\begin{figure}[t]
\begin{center}
\includegraphics{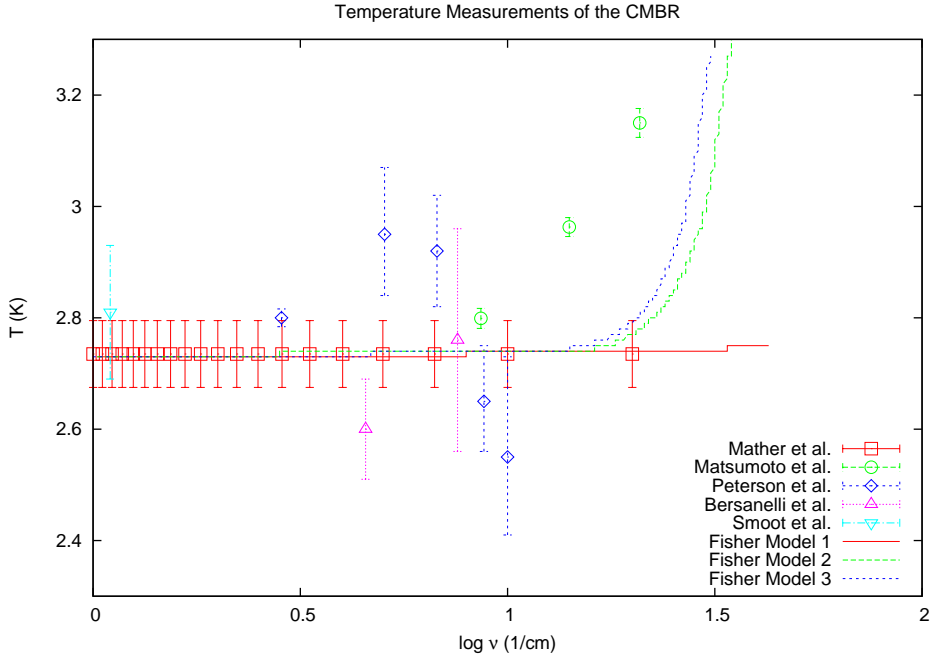}
\caption{Three different models for the distortion of the CMBR by inverse Compton scattering off of hot IGM electrons  is shown above. All three models have $\theta(z=2.5)=0.0025$. The models differ in the epoch at which the radiation from QSOs first ionized the IGM, these are Model 1 -- $z_{on} = 6$, Model 2 -- $z_{on} = 15$, and Model 3 -- $z_{on} = 20$. Also shown are previous temperature measurements of the CMBR by previous authors including COBE90. It is noted that all three models from this work are in agreement with the COBE90 measurements (including the uncertainty in the measurements).}
\label{cobe}
\end{center}
\end{figure}

The results are shown in Figure \ref{cobe}. Three different models are shown in Figure \ref{cobe} corresponding to different epochs when radiation from QSOs first ionized the IGM. These models are Model 1 -- $z_{on} = 6$, Model 2 -- $z_{on} = 15$, and Model 3 -- $z_{on} = 20$. The models start by adjusting the initial temperature of the CMBR so that the output temperature of the distorted CMBR is $T_{\gamma_0} = 2.735$ K in the range 1 cm $\geq \nu^{-1} \geq$ 0.05 cm, to correlate to COBE90. (This is done only to show that the results of COBE90 do not disallow a hot IGM.) Previous authors have adjusted the density and temperature of the IGM in order to model the cosmic x ray background and the highly distorted CMBR by Matsumoto \etal (1988). The initial scope of this work was to adjust the parameter $z_{on}$ in order to present a model of the IGM that also matched the results of previous measurements of the CMBR, discounting COBE90. The preliminary results of COBE90 have changed the path of this work. Whereas it has been boldly stated that COBE90 has shown there is not a hot (ionized) IGM, the results shown here show that COBE90 data is inconclusive as to the existence of the hot IGM. The COBE90 results do put limits on the parameter space of temperature and density of the IGM as evidence by Figure \ref{cobe}. A hot, thin IGM may still exist, but certainly not to the extent as to be the source of the cosmic x ray background as has been previously postulated, TW and FP.

\section{Summary}
Previous efforts, Fisher (preprint), calculated an effective recombination coefficient for optically thin hydrogen clouds. An evolutionary model of the diffuse UV background was then calculated and its effects on the stability of metal-free \ly cloud systems, because of radiation pressure, was modeled. From this model, limits were derived on the epoch of photoionization of the \ly clouds. Primordial nucleosynthesis and the observed mass of luminous matter in the Universe were used to determine the temperature and density of the IGM at the epoch $z=2.5$. This was used as a boundary condition to determine the thermal history of the IGM. Finally, the distortion of the CMBR because of this hot IGM was calculated and compared to previously published data. This mode was found to disagree with the distortion as measured by the Berkeley-Nagoya group, but was found to be consistent with the preliminary results of the Cosmic Background Explorer to high accuracy. This model of the IGM is not capable of producing the cosmic x ray background as measured.

The most interesting result of this work is its implication for the distortion of the CMBR. Much effort had been made to explain the distortion in the Wien tail of the CMBR, as measured prior to the launch of the COBE satellite. The original intent of this study was to proceed along the same path as previous authors, most notably TW, and determine if the then observed distortion could be attributed to inverse Compton scattering of the CMBR photons by hot IGM electrons. The hoped for end result was to compare the required density and temperature of the IGM to the observed conditions in the \ly forest cloud systems. Only after the model was complete, and the lack of a gross distortion similar to that published by the Berkeley-Nagoya group failed to appear, were the results of COBE made known. Whereas it has been stated that COBE disproves a hot IGM, this study indicates that the results, published to date, only dispel the previous results of the Berkeley-Nagoya group, and set a boundary, in temperature-density space, allowed for the IGM. A hot IGM may still exist and only after COBE extends the spectra to even shorter wavelengths, $\sim$ 0.01 cm, will the limits on the existence of the hot IGM proposed here be discounted or substantiated.

Several observations are needed to support the theoretical findings presented here, most notably the COBE spectra noted above. Other observations are needed to better define the limits on many of the physical characteristics of the \ly cloud systems, including size, temperature, and distribution in neutral hydrogen column density. All of these values were used to determine the temperature and density of the IGM. However, it is noted that drastic changes in the values of these parameters is not expected, and therefore the result of a hot IGM will not change. The results of extending the COBE spectra to shorter wavelengths is most anxiously awaited.

\acknowledgments
The author gratefully acknowledges J.A. Baldwin, E.R. Capriotti, G.J. Ferland, J.V. Villumsen and R.J. Weymann for helpful dscussions and comments on the manuscript. This work was supported in part by National Science Foundation Grant AST 000-00.

\end{document}